\documentclass[cameraready]{Interspeech}

\title{HALO: Half-Frame-Rate Adaptive Learnable Operator for Lightweight STFT-Based Speech Enhancement}

\author[affiliation={1,2}]{Jiadong}{Zhao}
\author[affiliation={1,2}]{Dahan}{Wang}
\author[affiliation={3}]{Yu}{Sun}
\author[affiliation={1,2}]{Leyan}{Yang}
\author[affiliation={1,2}]{Xiaobin}{Rong}
\author[affiliation={2}]{Shiruo}{Sun}
\author[affiliation={2}]{Yuxiang}{Hu}
\author[affiliation={1,2},correspondingauthor]{Jing}{Lu}

\address{
    $^1$ Key Laboratory of Modern Acoustics and Institute of Acoustics, Nanjing University, Nanjing, 210093, China \\
    $^2$ NJU-Horizon Intelligent Audio Lab, Horizon Robotics, Beijing 100094, China\\
    $^3$ R\&D Centre, Samsung Electronics (China), Nanjing 210019, China
}

\email{\{jiadong.zhao, dahan.wang\}@smail.nju.edu.cn, yuyu.sun@samsung.com, \{leyan.yang, xiaobin.rong\}@smail.nju.edu.cn, \{shiruo.sun, yuxiang.hu\}@horizon.cc, lujing@nju.edu.cn}


\usepackage{comment}

\begin{document}

\maketitle

\begin{abstract}
STFT-based speech enhancement typically adopts overlapping analysis frames. While overlap is essential for stable STFT processing, it makes adjacent frames highly correlated, causing redundant computation in lightweight models. We propose Half-frame-rate Adaptive Learnable Operator (HALO), a causal plug-in module that halves the internal frame rate without altering the STFT procedure. Broadly applicable to many lightweight models, HALO applies adaptive rate reduction before the backbone and restoration afterward, reconstructing the full-rate spectrum on the original STFT grid. Both reduction and restoration are implemented with lightweight dynamic convolutions. By halving the processed frame rate, HALO reduces backbone compute cost with no added algorithmic latency, freeing budget for channel widening. Experiments on the DNS3 dataset show consistent gains across diverse lightweight models under matched complexity, demonstrating the effectiveness of reducing overlap-induced redundancy.
\end{abstract}

\section{Introduction}
Speech enhancement (SE) aims to recover clean speech from adverse conditions such as background noise and reverberation, so as to improve speech quality and intelligibility \cite{SpeechEnhancement}. The DNN-based SE algorithms can be broadly categorized into time-frequency (T-F) domain approaches \cite{Tan2018CRN,Hu2020DCCRN,DPCRN,Schroeter2022DeepFilterNet2,Wang2023TFGridNet} and time-domain approaches \cite{Luo2019ConvTasNet,Luo2020DPRNN,Wang2021TSTNN}. Based on the short-time Fourier transform (STFT) representation, the T-F approaches effectively exploit the spectral structure to model speech and noise characteristics, and have long been a dominant paradigm in this field.

For resource-constrained scenarios on edge devices, a number of lightweight SE neural networks have emerged in recent years. These models reduce computational cost while maintaining competitive enhancement performance by designing efficient architectures. For instance, DPCRN \cite{DPCRN} combines convolutional recurrent network (CRN) \cite{Tan2018CRN} with dual-path recurrent neural network (DPRNN) \cite{Luo2020DPRNN} to better capture intra-frame spectral patterns and inter-frame dependencies. Building on DPCRN, GTCRN \cite{GTCRN} significantly simplifies the model with grouped operations and further leverages subband feature extraction and temporal recurrent attention modules to boost performance. Moreover, LiSenNet \cite{LiSenNet} introduces several effective modules and techniques, including sub-band convolution, convolutional gated linear unit, noise detection, and phase refinement, to strengthen lightweight modeling. More recently, UL-UNAS \cite{ULUNAS} utilizes neural architecture search to discover an optimal architecture based on the framework of GTCRN, achieving state-of-the-art results among ultra-lightweight models with the same or lower computational complexity.

Despite these advances in lightweight architecture design, they largely focus on reducing per-frame computation. Under traditional per-frame STFT processing, the per-second computational cost grows with the frame rate, which is determined by the STFT hop length. To mitigate boundary effects and enable smooth reconstruction, STFT commonly adopts overlapping analysis frames, and ISTFT reconstructs the waveform via overlap-add synthesis \cite{Allen1977UnifiedSTFT,Loizou2007SpeechEnhancement}. With 50\% or higher overlap, adjacent frames share many time-domain samples and are often strongly correlated, making the resulting STFT frame sequence temporally redundant \cite{Sturmel2011STFTMag}. This renders the overlap-induced redundancy a key bottleneck for improving compute efficiency beyond architecture-level design.

Prior overlap-related work, such as dual-window synthesis \cite{Wang2023STFTLowLatency} and multi-frame prediction \cite{Wang2022OverlappedFramePrediction,Bartolewska2023OFPDCCRN}, has been mainly developed for lower latency or improved reconstruction, and thus offers limited benefit for lowering the computational load. Moreover, these designs often depart from the traditional STFT procedure or the per-frame interface assumed by existing efficient backbones, requiring nontrivial redesign or adaptation. This motivates reducing compute by lowering the effective number of frames processed inside the backbone without modifying the STFT/ISTFT procedure.

In this work, we propose Half-frame-rate Adaptive Learnable Operator (HALO), a causal plug-in module that runs the enhancement backbone at a half frame rate via two adaptive learnable operators for rate reduction and restoration. Preserving the original STFT/ISTFT procedure, HALO is agnostic to the backbone architecture and can be readily integrated into existing lightweight STFT-based models. Specifically, HALO performs rate reduction before the backbone and rate restoration afterward to recover the full-rate complex spectrum on the original STFT grid. To preserve speech details and stability under half-frame-rate backbone processing, we design adaptive learnable rate-reduction/restoration operators based on lightweight dynamic convolutions \cite{Yang2019CondConv,Chen2020DynamicConv,Wang2021Per-pixel,AdaptCRN}, enabling fusion and restoration conditioned on local T-F features. By shortening the sequence processed by the backbone, HALO reduces average backbone multiply-accumulate operations per second (MAC/s), freeing compute for channel widening under comparable complexity. We apply HALO to multiple lightweight backbones and demonstrate consistent improvements over their original counterparts on the DNS3 dataset.  The audio samples are available at \url{https://github.com/dddaniel-z/HALO/}.

\section{Methodology}

\subsection{Problem formulation}

We consider monaural speech enhancement in the T-F domain. In the time domain, the observed signal $x(t)$ is modeled as the clean speech $s(t)$ corrupted by additive distortion $n(t)$.
Applying the STFT yields the complex spectrograms $X$, $S$, and $N$,
where we represent $X = \left[ \Re\{X\}; \Im\{X\} \right] \in \mathbb{R}^{2 \times T \times F}$ with real and imaginary parts along the first dimension,
and $T$ and $F$ denote the numbers of time frames and frequency bins. Under the additive
assumption in the T-F domain, we have
\begin{align}
X(t, f) = S(t, f) + N(t, f).
\end{align}
Our objective is to estimate an enhanced spectrum $\hat{S}$ from $X$, and the enhanced waveform $\hat{s}$ can be reconstructed by the ISTFT of $\hat{S}$.

\subsection{HALO overview}
\begin{figure}[htb]
	\hfill
	\includegraphics[width=0.9\linewidth]{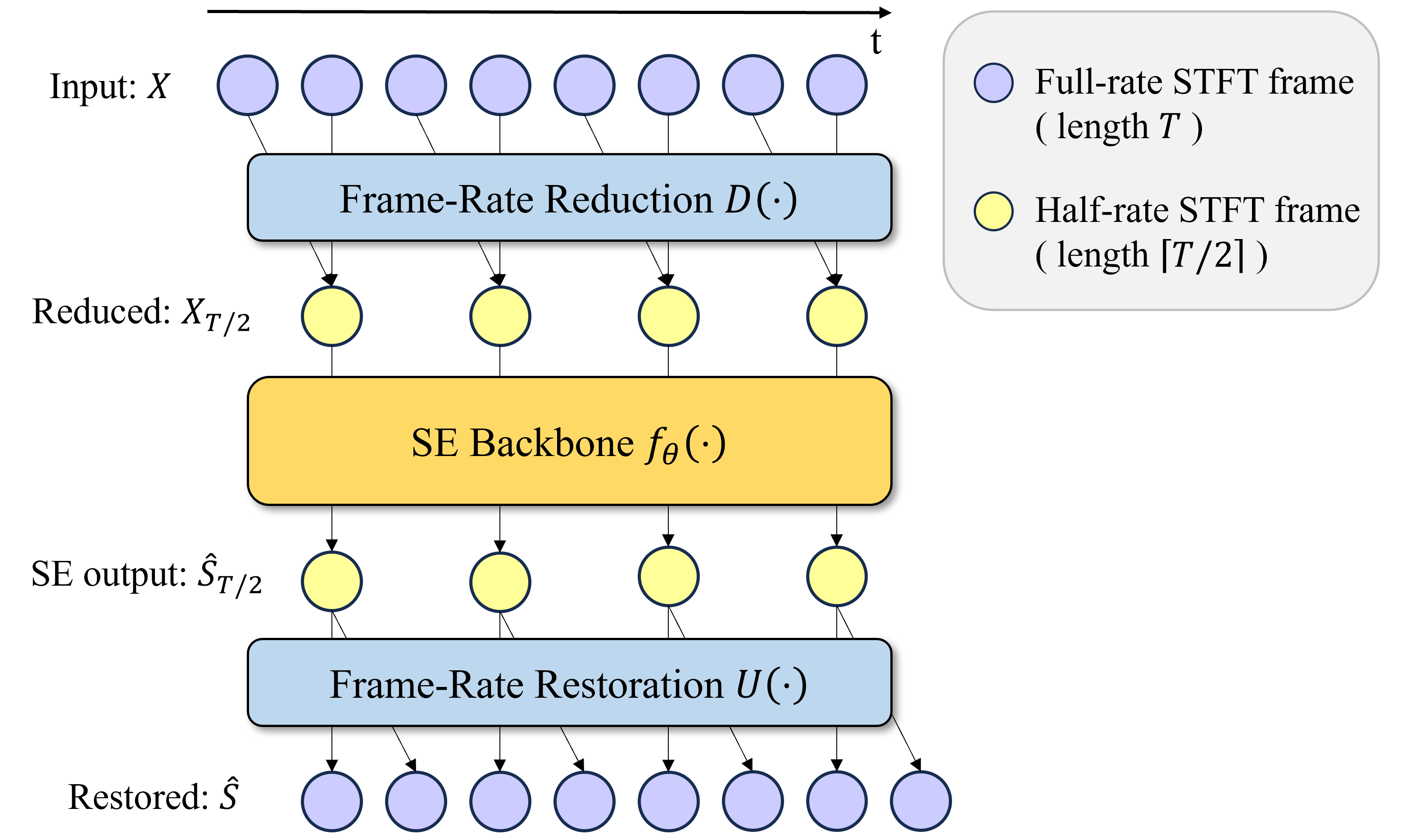}
	\caption{Overall framework of HALO}
    \label{Fig1}
\end{figure}

HALO is a plug-in module that reduces the internal frame rate of a backbone while preserving the original output resolution on the STFT grid. It consists of two adaptive learnable operators, $D(\cdot)$ for frame-rate reduction and $U(\cdot)$ for frame-rate restoration, as shown in Fig.~\ref{Fig1}. 

Specifically, we first apply a temporal rate reduction operator $D(\cdot)$ to compress the input sequence:
\begin{align}
X_{T/2} = D(X) \in \mathbb{R}^{2 \times \lceil T/2 \rceil \times F}.
\end{align}

The backbone then operates on the shortened sequence to produce half-frame-rate outputs:
\begin{align}
\hat{S}_{T/2} = f_\theta(X_{T/2}) \in \mathbb{R}^{2 \times \lceil T/2 \rceil \times F}.
\end{align}
Here, $f_\theta(\cdot)$ denotes the STFT-based enhancement backbone parameterized by $\theta$.

Finally, a restoration operator $U(\cdot)$ expands each half-rate frame into two adjacent frames, restoring the full-rate sequence on the original STFT grid:
\begin{align}
\hat{S} = U(\hat{S}_{T/2}) \in \mathbb{R}^{2 \times T \times F}.
\end{align}

Concretely, each reduced frame fuses the current frame with its immediate predecessor on the original grid, and the restoration predicts the current frame together with its immediate successor. Therefore, HALO can be inserted without requiring any future STFT frames, preserving the backbone's per-frame interface and algorithmic latency.

\subsection{Frame-rate reduction operator}
\begin{figure}[htb]
	\centering
	\includegraphics[width=0.9\linewidth]{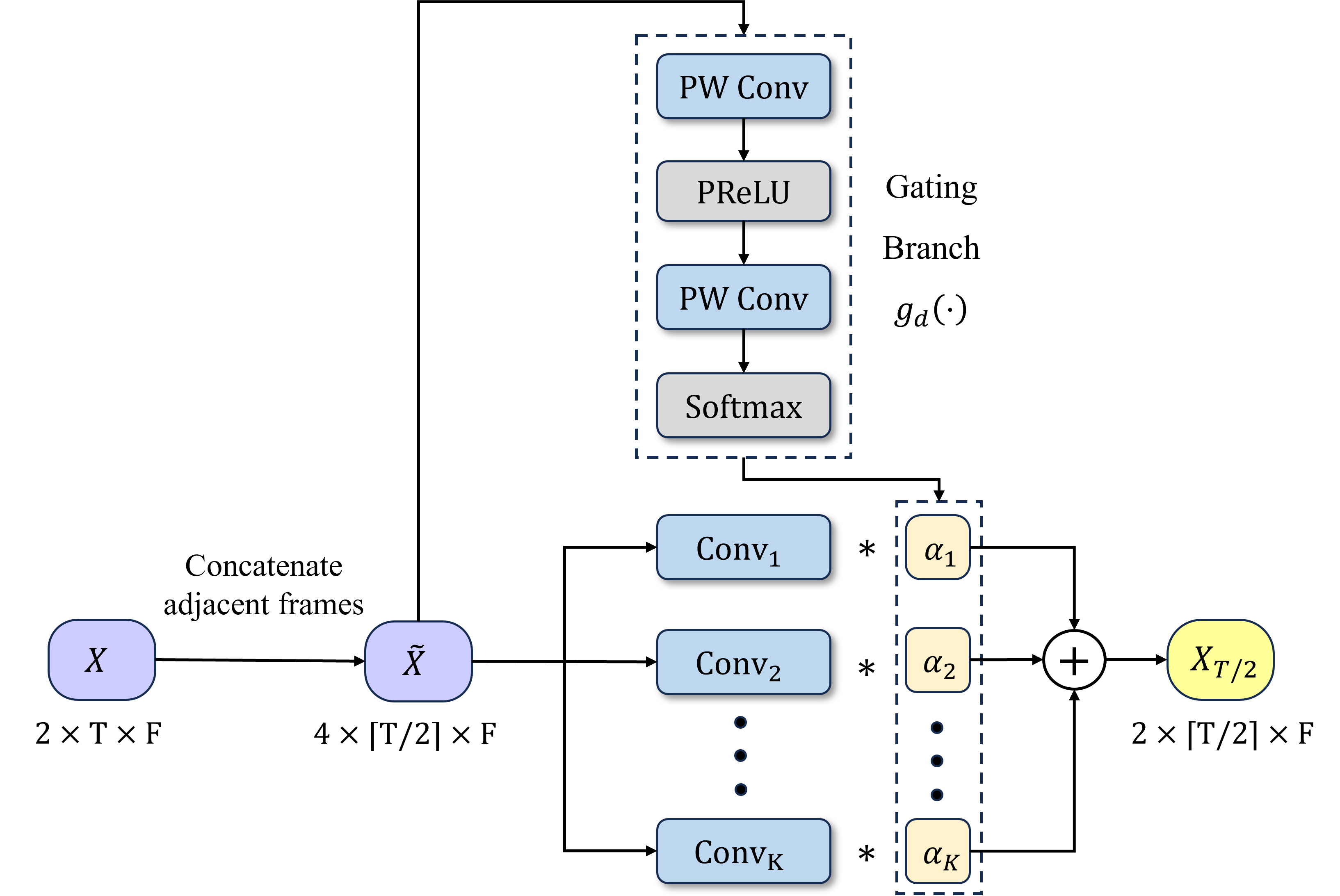}
	\caption{Adaptive learnable rate-reduction operator}
    \label{Fig2}
\end{figure}

The rate reduction operator $D(\cdot)$ is realized by an adaptive learnable module, which integrates the information from adjacent temporal frames instead of simply discarding every other STFT frame. The overview of the operator is depicted in Fig.~\ref{Fig2}. Let $X(:, t, f) \in \mathbb{R}^{2}$
denote the complex feature at frame index $t$ and frequency bin $f$. With $L = \lceil T/2 \rceil$, for each reduced-time index $l = 0, \dots, L-1$, we form a two-frame paired feature $\tilde{X}(:, l, f)$ by concatenating adjacent frames at each frequency bin:
\begin{align}
\tilde{X}(:, l, f) = \text{cat}(X(:, 2l-1, f), X(:, 2l, f)) \in \mathbb{R}^4,
\end{align}
where $X(:, -1, f)$ is set to an all-zero vector for all $f$ to handle the boundary case. $\tilde{X}(:, l, f)$ is then mapped to a half-frame-rate representation $X_{T/2}(:, l, f) \in \mathbb{R}^2$ using a dynamic convolution with lightweight gating branch. 

Specifically, we maintain a kernel bank $\{W_k\}_{k=1}^K$, with $W_k \in \mathbb{R}^{2 \times 4}$, and predict T-F-dependent mixture weights $\alpha_k(l, f)$ via a lightweight gating function $g_d(\cdot)$:
\begin{gather}
\alpha(l, f) = g_d(\tilde{X}(:, l, f)), \\ \sum_{k=1}^K \alpha_k(l, f) = 1,\ \alpha_k(l, f) \geq 0.
\end{gather}
Here $g_d(\cdot)$ denotes a lightweight gating network consisting of two point-wise (PW) convolutions and a PReLU nonlinearity, followed by a softmax over the kernel index. Thus, $g_d(\cdot)$ outputs a set of $K$ normalized mixture weights at each T-F bin. 
\begin{table*}[htbp]
\renewcommand{\arraystretch}{1}
	\centering
    \caption{Ablation study results on DNS3 test set.}
    {\fontsize{8}{9.5}\selectfont
	\begin{tabular}{c|cc|cccccc}
		\hline
		             & Para. (k) & MAC/s (M) & PESQ  & ESTOI & SI-SNR  & OVRL & SIG & BAK   \\ \hline
		Noisy        &           &            & 1.406 & 0.669 & 5.610  & 1.628 & 2.049 & 1.855 \\ 
		GTCRN (baseline)        & 23.67     & 33.83      & 2.101 & 0.754 & 11.390 & 2.629 & 2.981 & 3.809 \\ \hline
		no-overlap STFT & 47.30     & 31.84      & 1.783 & 0.722 & 10.960 & 2.512 & 2.859 & 3.728 \\ \hline
  		D-FixedRed + U-FixedRest   & 47.40     & 32.02      & 2.118 & 0.758 & 11.620 & 2.663 & 3.003 & \textbf{3.853} \\
    	D-Decimate + U-FixedRest  & 47.32     & 32.70      & 2.104 & 0.755 & 11.510 & 2.600 & 2.936 & 3.817 \\
		D-Decimate + U-Duplicate & 47.30     & 32.60      & 2.086 & 0.752 & 11.400 & 2.568 & 2.929 & 3.794 \\ \hline
        HALO (w/o channel widening)     & 24.07   &   22.05
        & 2.093	&0.754	&11.430	&2.625	&2.970	&3.826 \\
		HALO (w/ channel widening)        & 46.87     & 32.85      & \textbf{2.198} & \textbf{0.769} & \textbf{11.900} & \textbf{2.673} &\textbf{3.015} & 3.843 \\ \hline
	\end{tabular}
 }
	\label{tab:Table1}
\end{table*}

We first compute $K$ basis outputs using the kernel bank, and then obtain the reduced feature by a weighted sum using $\alpha_k(l,f)$:
\begin{gather}
{X_{T/2,k}}(:, l, f) = W_k\tilde{X}(:, l, f), \\
X_{T/2}(:, l, f) = \sum_{k=1}^K \alpha_k(l, f) {X_{T/2,k}}(:, l, f).
\end{gather}

This implementation is equivalent to mixing the kernels before convolution, but is more implementation-friendly under time-varying adaptive weights, with similar MAC/s under our channel configuration. Because the fusion weights are conditioned on the local T-F feature, the frame-rate reduction operator can adaptively combine the two adjacent frames, either emphasizing one frame or forming a mixture of both. This flexibility enables temporal compression while retaining information that is important for reconstructing rapidly varying speech components.

\subsection{Frame-rate restoration operator}
Given the half-frame-rate backbone output $\hat{S}_{T/2} \in \mathbb{R}^{2 \times L \times F}$, the restoration operator $U(\cdot)$ reconstructs the full-rate STFT sequence $\hat{S} \in \mathbb{R}^{2 \times T \times F}$ on the original grid. For each reduced-time index $l$, we synthesize two consecutive frames corresponding to original indices $2l$ and $2l+1$ with a dynamic convolution. The restoration operator serves as a structural counterpart of the reduction, using the same kernel-bank and gated design.

Specifically, for each bin $(l,f)$, we maintain a kernel bank $\{V_k\}_{k=1}^K$ with $V_k \in \mathbb{R}^{4 \times 2}$, and predict T-F-dependent mixture weights $\beta_k(l,f)$ via the gating function $g_u(\cdot)$, which shares the same architecture as $g_d(\cdot)$:
\begin{gather}
\beta(l,f) = g_u(\hat{S}_{T/2}(:,l,f)), \\
\sum_{k=1}^K \beta_k(l,f) = 1,\ \beta_k(l, f) \geq 0
\end{gather}

We first compute $K$ basis restoration outputs using the kernel bank, and then form the restored two-frame pair by a weighted sum using $\beta_k(l,f)$:
\begin{gather}
\tilde{S}_k(:, l, f) = V_k \hat{S}_{T/2}(:, l, f),\\
\tilde{S}(:, l, f) = \sum_{k=1}^K \beta_k(l, f) \tilde{S}_k(:, l, f) \in \mathbb{R}^4.
\end{gather}
We then split $\tilde{S}(:, l, f)$ into two complex frames:
\begin{gather}
\hat{S}(:, 2l, f)   = \tilde{S}(1:2, l, f), \\
\hat{S}(:, 2l+1, f) = \tilde{S}(3:4, l, f).
\end{gather}
When $T$ is odd, we truncate the last synthesized frame to match the original sequence length. Here, $\hat{S}(:,2l,f)$ and $\hat{S}(:,2l+1,f)$ correspond to the current frame and the immediately subsequent frame on the original STFT grid.
 
Notably, the reduction at index $l$ depends only on $X(:, 2l-1, :)$ and $ X(:, 2l, :)$, while the restoration produces $\hat{S}(:, 2l, :)$ and $\hat{S}(:, 2l+1, :)$ solely from $\hat{S}_{T/2}(:,l,:)$ without accessing any future input. This design introduces no additional lookahead and does not increase algorithmic latency.

\section{Experiments}
\subsection{Datasets}
To evaluate the performance of our proposed framework, we conduct experiments on the 3rd Deep Noise Suppression (DNS3) dataset \cite{2021DNSChallenge}, which contains a wide range of clean sets, noise sets, and room impulse responses (RIRs). Additionally, we also include the Mandarin corpus from DiDiSpeech \cite{2021DiDiSpeech}. For data generation, the clean speech is convolved with a randomly selected RIR, and then mixed with randomly selected noise clips with the signal-to-noise ratio (SNR) ranging from $-5$ to $15$ dB. A total of 72\,000 pairs of 10-second noisy-clean data are selected for training, while 840 and 800 pairs are generated for validation and testing, respectively. All the utterances are sampled at \SI{16}{\kilo\hertz}. 

\subsection{Implementation details}
STFT is computed using a \SI{32}{\milli\second} square-root Hann window, a \SI{16}{\milli\second} hop length (50\% overlap), and an FFT length of 512, unless noted. For dynamic convolution, the number of candidate kernels is set to 5, with 8 hidden channels in the gating branch.

The models are trained by Adam Optimizer \cite{Adam} with an initial learning rate of 0.001. The learning rate will be halved if the validation loss does not decrease for 10 consecutive epochs. During training, we use a batch size of 8 and adopt the same training loss function as in GTCRN \cite{GTCRN} for all experiments.

The model complexity is measured by the number of parameters and MAC/s. For controlled comparisons, we widen the backbone channel width when inserting HALO so that the resulting model operates at similar MAC/s to the corresponding baseline. This isolates the effect of HALO under comparable computational cost. We match MAC/s rather than parameter count because parameter count can be a misleading proxy due to varying parameter efficiency of different architectures \cite{Zhang2024BeyondPlateaus}. Moreover, for lightweight models, parameters mainly reflect memory footprint, whereas real-time edge deployment is typically compute bottlenecked \cite{ULUNAS}.

\begin{table*}[htbp]
\renewcommand{\arraystretch}{1}
	\centering
     \caption{Experiments across diverse backbones on DNS3 test set.}
     {\fontsize{8}{10}\selectfont
	\begin{tabular}{l|ll|llllll}
		\hline
		Model                 & Para. (k) & MAC/s (M) & PESQ  & ESTOI & SI-SNR  & OVRL  & SIG   & BAK   \\ \hline
		Noisy                 &          &          & 1.406 & 0.669 & 5.610  & 1.628 & 2.049 & 1.855 \\ \hline
        GTCRN-50\% overlap    & 23.67     & 33.83      & 2.101 & 0.754 & 11.390 & 2.629 & 2.981 & 3.809 \\ 
        GTCRN-50\% overlap + HALO         & 46.87     & 32.85      & 2.198 & 0.769 & 11.900 & 2.673 & 3.015 & 3.843 \\ \hline
  	   GTCRN-75\% overlap              & 23.67    & 67.66    & 2.121 & 0.756 & 11.370 & 2.639 & 2.977 & 3.846 \\
		GTCRN-75\% overlap + HALO         & 46.87    & 65.49    & 2.237 & 0.772 & 11.990 & 2.674 & 3.011 & 3.857 \\ \hline
		DPCRN-ultralight      & 27.92    & 31.80    & 2.025 & 0.750 & 11.070 & 2.597 & 2.943 & 3.797 \\
		DPCRN-ultralight + HALO & 55.03    & 31.34    & 2.212 & 0.771 & 11.920 & 2.648 & 2.990 & 3.826 \\ \hline
		DPCRN-light           & 57.02    & 59.84    & 2.146 & 0.764 & 11.550 & 2.631 & 2.975 & 3.810 \\
		DPCRN-light + HALO      & 115.95   & 61.48    & 2.247 & 0.778 & 12.100 & 2.657 & 2.995 & 3.842 \\ \hline
		DPCRN-middle          & 207.22   & 225.11   & 2.340 & 0.788 & 12.350 & 2.713 & 3.048 & 3.865 \\
		DPCRN-middle + HALO     & 419.80   & 223.45   & 2.402 & 0.796 & 12.670 & 2.708 & 3.046 & 3.859 \\ \hline
		DPCRN-large           & 787.15   & 872.09   & 2.522 & 0.807 & 13.130 & 2.777 & 3.115 & 3.887 \\
		DPCRN-large + HALO      & 1440.00  & 869.86   & 2.536 & 0.809 & 13.150 & 2.766 & 3.105 & 3.882 \\ \hline
		LiSenNet              & 36.78    & 55.77    & 2.177 & 0.762 & 11.760 & 2.681 & 3.029 & 3.837 \\
		LiSenNet + HALO         & 64.52    & 55.76    & 2.275 & 0.778 & 12.390 & 2.703 & 3.047 & 3.847 \\ \hline
		UL-UNAS               & 168.98   & 33.61    & 2.245 & 0.773 & 12.100 & 2.681 & 3.011 & 3.859 \\
		UL-UNAS + HALO          & 205.16   & 31.26    & 2.261 & 0.777 & 12.240 & 2.684 & 3.027 & 3.848 \\ \hline
	\end{tabular}
    }
	\label{tab:Table2}
\end{table*}
\subsection{Ablation study}
We conduct ablation experiments on GTCRN to verify the efficacy of HALO. The evaluation is conducted on the test set using metrics including perceptual evaluation of speech quality (PESQ) \cite{PESQ}, extended short-time objective intelligibility (ESTOI) \cite{ESTOI}, scale-invariant signal-to-noise ratio (SI-SNR) \cite{SISNR} and DNSMOS P.835 \cite{DNSMOSP835}. The ablation test results are presented in Table~\ref{tab:Table1}.

As a reference, we remove STFT overlap by setting the hop size equal to the window length and using the rectangular window (no-overlap STFT). This change causes a pronounced degradation across all metrics, confirming that overlap cannot be simply discarded without hurting enhancement quality. We then conduct the remaining ablations under the original STFT setting. Specifically, we progressively remove adaptive gating, learnable reduction/restoration, and channel widening to disentangle their individual contributions.

We first disable adaptive gating in both reduction and restoration, yielding learnable but fixed-kernel single-convolution operators (D-FixedRed + U-FixedRest). This degrades performance relative to HALO and highlights the benefit of adaptive gating. Replacing the fixed-kernel reduction with simple decimation by dropping every other STFT frame (D-Decimate + U-FixedRest) further weakens performance, indicating that a learnable reduction is important. Substituting the fixed-kernel restoration with duplicating each half-rate output to two adjacent frames yields the simplest variant (D-Decimate + U-Duplicate) and causes another drop, showing that learnable restoration is also necessary. 

Table~\ref{tab:Table1} also shows that HALO without channel widening substantially reduces computational complexity (22.05M vs. 33.83M MAC/s) while maintaining near-baseline performance. In the cost-matched setting with channel widening, HALO achieves the best overall results, indicating that exploiting overlap-induced redundancy with adaptive learnable operators is crucial for preserving enhancement quality under frame-rate compression. Compared with the GTCRN baseline, HALO (w/ channel widening) improves PESQ by 0.1 and SI-SNR by 0.5 dB, and also boosts DNSMOS-OVRL by 0.04, with the algorithmic latency unchanged.

\subsection{Comparison across different backbones}
Table~\ref{tab:Table2} reports the plug-in results of HALO on several lightweight STFT-based enhancement backbones, including DPCRN \cite{DPCRN} of different sizes, GTCRN \cite{GTCRN}, LiSenNet \cite{LiSenNet} and UL-UNAS \cite{ULUNAS}. HALO consistently improves objective metrics across all tested architectures under comparable computational cost, demonstrating that overlap-induced temporal redundancy is a common efficiency bottleneck in these lightweight backbones. Notably, to test heavier STFT redundancy, we also evaluate GTCRN under a 75\% overlap STFT setting (\SI{32}{\milli\second} window, \SI{8}{\milli\second} hop), in which case HALO reduces the internal frame overlap ratio to 50\% and still delivers performance gains after channel widening. These results demonstrate that HALO addresses a bottleneck distinct from architectural slimming, providing consistent improvements on top of existing lightweight designs.

The DPCRN family illustrates that the magnitude of improvement depends on the backbone capacity. When the backbone has limited capacity, the model is forced to spend substantial computation on temporally redundant overlapping frames. By reducing the internal frame rate, HALO mitigates redundancy and concentrates computation on more effective modules, leading to a noticeable improvement. As model size increases, the gain brought by HALO becomes smaller, as the additional channels become less effective once the capacity is already large. This trend is also observed in UL-UNAS, whose backbone is already highly optimized. Consequently, reallocating the saved compute simply through channel widening yields diminishing returns, leading to modest gains. It should be noted that we focus on reducing overlap-induced redundancy in this work, rather than exhaustively optimizing compute reallocation strategies. Importantly, the compute saved by reducing overlap-induced redundancy still provides headroom for incorporating stronger modules or more efficient architectural designs.

\subsection{Discussion}
HALO reduces the average computational cost by halving the internal sequence processed by the backbone, and in our cost-matched setting the saved temporal budget is reallocated to channel widening. However, HALO does not reduce the peak per-step computation, since the frame-rate restoration operator generates two adjacent frames within the same inference step. Designing peak-aware variants of HALO, including peak-constrained compute reallocation and scheduling strategies for streaming deployment, is left for future work.

\section{Conclusion}
We propose HALO, a causal plug-in module for lightweight STFT-based speech enhancement that reduces the overlap-induced temporal redundancy. By introducing adaptive learnable operators for frame-rate reduction and restoration, HALO can be integrated into existing backbones and halve the internal frame rate. Ablation results on DNS3 support the effectiveness of the proposed design, and evaluations across multiple lightweight backbones show consistent gains under comparable computational complexity.

\section{Acknowledgments}
This work is supported by National Natural Science Foundation of China (Grants No.12274221) and the AI \& AI for Science Project of Nanjing University.

\section{Generative AI Use Disclosure}
Generative AI tools were used only for language editing and polishing. They were not used to generate research ideas, experimental designs, results, or core technical content. The authors verified all statements, code, and references and take full responsibility for the manuscript.

\bibliographystyle{IEEEtran}
\bibliography{mybib}

\end{document}